\documentclass[12pt]{article}
\usepackage{epsf}

\renewcommand{\thefootnote}{\#\arabic{footnote}}
\newcommand{\gtrsim}{ \mathop{}_{\textstyle \sim}^{\textstyle >} }
\newcommand{\lesssim}{ \mathop{}_{\textstyle \sim}^{\textstyle <} }

\begin{document}

\renewcommand{\thefootnote}{\fnsymbol{footnote}}
\setcounter{footnote}{0}
\begin{titlepage}

\def\thefootnote{\fnsymbol{footnote}}

\begin{center}

\hfill TU-669\\
\hfill UCB-PTH-02/49\\
\hfill LBNL-51702\\
\hfill hep-ph/0211019\\
\hfill November, 2002\\

\vskip .5in

{\Large \bf

CMB Anisotropy from Baryogenesis by a Scalar Field

}

\vskip .45in

{\large
Takeo Moroi$^{(a)}$ and Hitoshi Murayama$^{(b,c)}$
}

\vskip .45in

{\em
$^{(a)}$Department of Physics, Tohoku University, Sendai 980-8578, Japan
}

\vskip .1in

{\em
$^{(b)}$Department of Physics, University of California, Berkeley, CA
94720
}

\vskip .1in

{\em
$^{(c)}$Theory Group, Lawrence Berkeley National Laboratory, Berkeley, 
CA 94720
}

\end{center}

\vskip .4in

\begin{abstract}

    We study the cosmic microwave background (CMB) anisotropy in the
    scenario where the baryon asymmetry of the universe is generated
    from a condensation of a scalar field.  In such a scenario, the
    scalar condensation may acquire fluctuation during the inflation
    which becomes a new source of the cosmic density perturbations.
    In particular, the primordial fluctuation of the scalar
    condensation may induce correlated mixture of the adiabatic and
    isocurvature fluctuations.  If the scalar condensation decays
    before it completely dominates the universe, the CMB angular power
    spectrum may significantly deviate from the conventional adiabatic
    result.  Such a deviation may be observed in the on-going MAP
    experiment.

\end{abstract}

\end{titlepage}

\renewcommand{\thepage}{\arabic{page}}
\setcounter{page}{1}
\renewcommand{\thefootnote}{\#\arabic{footnote}}
\setcounter{footnote}{0}

\renewcommand{\theequation}{\thesection.\arabic{equation}}

\section{Introduction}
\label{sec:introduction}
\setcounter{equation}{0}

In modern particle cosmology, one of the most important issues is to
understand the origin of the baryon asymmetry of the universe.  In
particular, assuming inflation \cite{PRD23-347} as a solution to the
horizon, flatness, and other cosmological problems and as a seed for
density fluctuations for later structure formation, we are obliged to
adopt scenarios where the baryon asymmetry is generated after the
reheating.

One of the main branches of baryogenesis is to use a primordial
condensation of a scalar field as a source of the baryon-number
asymmetry of the universe.  In such a class of scenarios, there exists
a primordial condensation of a scalar field and it decays at a later
stage of the evolution of the universe.  Then, baryon asymmetry is
generated by the decay of the scalar field.  If such a scalar field
exists, one of the Sakharov's three conditions for baryogenesis, i.e.,
the out-of-equilibrium condition, is easily satisfied by assuming
non-vanishing initial amplitude of the scalar field.  In particular,
in cosmological scenarios based on supersymmetric models, there exist
various possible candidates of such a scalar field and many efforts
have been made to study such scenarios.

Among various possibilities, probably one of the most well-motivated
candidates of such scalar fields is the superpartner of the
right-handed neutrino, the right-handed sneutrino.  Assuming the
seesaw mechanism \cite{seesaw} as a origin of very tiny neutrino
masses which are suggested from the neutrino-oscillation experiments
\cite{nu-osc},\footnote
{Even for Dirac neutrinos, an analogue of the seesaw mechanism is
possible \cite{PRD64-115011,PRD64-053005} and the leptogenesis
is possible \cite{PRL84-4039} via the decay of a scalar condensate
\cite{hph0206177}.}
the right-handed sneutrinos inevitably exist in the
supersymmetric models.  Such a right-handed sneutrino can be a source
of the baryon-number asymmetry \cite{PLB322-349,PRD65-043512}; if one
of the right-handed neutrinos has non-vanishing amplitude in the early
universe, its decay may generate lepton-number asymmetry which is
converted to the baryon-number asymmetry via the sphaleron interaction
\cite{PLB155-36}.  (In fact, this is a supersymmetric extension of the
leptogenesis scenario proposed by Fukugita and Yanagida
\cite{PLB174-45}.)

In addition, Affleck-Dine mechanism for baryogenesis \cite{NPB249-361}
is another possibility.  In the Affleck-Dine scenario, using the
baryon-number violating operator as a source, non-vanishing baryon
number is generated while a scalar partner of quarks, called
Affleck-Dine field, is oscillating.

Since the baryogenesis by a scalar condensation is attractive and
well-motivated in particular in supersymmetric models, it is important
to consider how the scenario can be experimentally tested.  As we will
discuss below, one possible effect of such a scalar field is on the
cosmological density perturbations; if a condensation of the scalar
field in the early universe is the source of the present baryon
asymmetry, some signal may be imprinted in the cosmic microwave
background (CMB) since the significant fraction of the CMB radiation
we observe today may also originate in the scalar-field condensation.
In particular, the amplitude of the scalar field may acquire sizable
fluctuation during the inflation and such a fluctuation affects the
CMB anisotropy.

Thus, in this paper, we study the cosmic density perturbations in the
scenario of the baryogenesis by a scalar field.  Although our
conclusions are quite general to a large class of scenarios, we mostly
concentrate on the scenario of the sneutrino-induced leptogenesis
\cite{PLB322-349,PRD65-043512} to make our discussion clearer.  In
particular, we study effects of the primordial fluctuation of the
scalar field on the CMB anisotropy.  If the primordial fluctuation
exists in the amplitude of the scalar field, correlated mixture of the
adiabatic and isocurvature fluctuations may be generated which may
significantly affect the CMB angular power spectrum.  With the precise
measurement of the CMB angular power spectrum by the MAP experiment,
some signal of the scalar-field-induced leptogenesis may be observed.

\section{The Leptogenesis Scenario}
\label{sec:scenario}
\setcounter{equation}{0}

Let us start by introducing the scenario we consider.  For our
argument, there are two scalar fields which play important roles; one
is the inflaton field $\chi$ and the other is the (lightest)
right-handed sneutrino.  In the following, we denote the {\sl
lightest} right-handed scalar neutrino as $\tilde{N}$.  We assume the
inflation so that the universe starts with a de Sitter epoch when the
universe is dominated by the potential energy of the inflaton field
$\chi$.  During the inflation, the right-handed sneutrino is assumed
to have a non-vanishing amplitude $\tilde{N}_{\rm init}$.  We treat
$\tilde{N}_{\rm init}$ as a free parameter in the following
discussion.  After the inflation, the inflaton field starts to
oscillate and then decays.  The right-handed neutrino also starts to
oscillate when the expansion rate of the universe becomes comparable
to the Majorana mass $M_N$.  (We assume that the expansion rate during
the inflation is larger than $M_N$ so that $\tilde{N}$ starts to
oscillate after the inflation.)  Then, the right-handed sneutrino
decays when $H\sim\Gamma_N$, where $\Gamma_N$ is the decay rate of
$\tilde{N}$.

One important point in this scenario is that the CMB radiation we
observe today has two origins; the inflaton field and the right-handed
sneutrino.  This is because the decays of the inflaton and the
right-handed sneutrino both convert the energy densities stored in the
scalar condensation into that of radiation.  For our study, it is
convenient to distinguish the radiation originating in the inflaton
field from that originating in the right-handed sneutrino.  We denote
them as $\gamma_\chi$ and $\gamma_{\tilde{N}}$, respectively.\footnote
{In fact, we cannot neglect the momentum transfer between these
photons and they cannot be defined separately.  In particular,
velocity perturbations of these photons should be the same since the
mean free path of the photon is much shorter than the horizon scale.
If we only discuss the behavior of $\delta_{\gamma_X}$ for the
super-horizon modes, however, the following discussions are valid since
the velocity perturbations are suppressed by the factor $k\tau$
relative to the density perturbations.  In a rigorous sense,
$\gamma_\chi$ and $\gamma_{\tilde{N}}$ should be understood as
representatives of the components produced from the decay products of
the inflaton field and $\tilde{N}$, respectively.}
Energy fraction of the radiation from $\tilde{N}$ depends on the
initial amplitude $\tilde{N}_{\rm init}$ as well as the decay rate of
the inflaton $\chi$ (denoted as $\Gamma_\chi$), decay rate of
$\tilde{N}$ and the mass of the right-handed (s)neutrino.

If $\Gamma_\chi\lesssim\Gamma_N$, $\tilde{N}$ decays when the universe
is dominated by the inflaton condensation.  In particular, when
$\Gamma_N\lesssim H\lesssim M_N$, energy densities of $\tilde{N}$ and
$\chi$ are related as\footnote
{When the initial amplitude of $\tilde{N}$ is larger than $M_*$,
inflation is caused by the energy density of $\tilde{N}$.  If this
happens, all the components generated from $\chi$ is washed out and
$f_{\gamma_{\tilde{N}}}\rightarrow 1$.  If the $e$-folding number of
the secondary inflation caused by $\tilde{N}$ is small enough, the
following discussions are unchanged since the density fluctuation for
the scale corresponding to the CMB anisotropy we will discuss is
generated from the primary inflation caused by $\chi$.  
If the initial amplitude is much larger than $M_*$, only the energy
density from the decay of the right-handed neutrino is relevant and
the perturbation becomes purely adiabatic \cite{PRL70-1912}.}
\begin{eqnarray}
\rho_{\tilde{N}} / \rho_\chi \sim \tilde{N}_{\rm init}^2 / M_*^2~:~~~
\Gamma_N \lesssim H\lesssim M_N.
\end{eqnarray}
(Here and hereafter, $\rho_X$ denotes the energy density of the
component $X$.)  Using this relation, we can evaluate the energy
density of the right-handed sneutrino at the time of its decay; when
$H\sim\Gamma_N$, $\rho_\chi\sim\Gamma_N^2M_*^2$ and hence
$\rho_{\tilde{N}}\sim\Gamma_N^2\tilde{N}_{\rm init}^2$.  After the
decay, energy density stored in the sneutrino condensation is
converted to that of radiation, so the universe is filled with the
radiation (denoted as $\gamma_N$) as well as the inflaton oscillation.
Using the relations $\rho_\chi\propto a^{-3}$ and
$\rho_{\tilde{N}}\propto a^{-4}$, we obtain the energy densities of
each components at $H\sim\Gamma_\chi$ as
$\rho_\chi\sim\Gamma_\chi^2M_*^2$ and
$\rho_{\gamma_{\tilde{N}}}\sim\rho_\chi
(\Gamma_\chi/\Gamma_N)^{2/3}(\tilde{N}_{\rm init}/M_*)^2$.  Thus,
after the decay of the inflaton field, the energy fraction of the
radiation from the right-handed sneutrino is estimated as
\begin{eqnarray}
f_{\gamma_{\tilde{N}}} \sim
\frac
{(\Gamma_\chi/\Gamma_N)^{2/3}(\tilde{N}_{\rm init}/M_*)^2}
{1+(\Gamma_\chi/\Gamma_N)^{2/3}
(\tilde{N}_{\rm init}/M_*)^2}
~~~:~~~\Gamma_\chi \lesssim\Gamma_N.
\label{f(Gn>Gx)}
\end{eqnarray}
Here, $f_{\gamma_X}$ is the energy fraction of $\gamma_X$ after the
decays of the inflaton and $\tilde{N}$, and $f_{\gamma_{\chi}}
+f_{\gamma_{\tilde{N}}}=1$.  (Here and hereafter,
$\gamma_X=\gamma_{\tilde{N}}$ or $\gamma_\chi$.)

If $\Gamma_\chi \gtrsim\Gamma_N$, on the contrary, $\tilde{N}$
decays after the inflaton decay.  In this case, it is convenient to
define the following quantity:
\begin{eqnarray}
\tilde{N}_{\rm eq} \sim 
\left\{
\begin{array}{ll}
(\Gamma_N / M_N)^{1/4} M_* & 
~~~:~~~ M_N < \Gamma_\chi \\
(\Gamma_N /\Gamma_\chi)^{1/4} M_* &
~~~:~~~ M_N > \Gamma_\chi
\end{array}
\right. .
\end{eqnarray}
If $\tilde{N}_{\rm init}\sim \tilde{N}_{\rm eq}$,
$\rho_{\gamma_\chi}\sim\rho_{\tilde{N}}$ is realized when
$H\sim\Gamma_{\tilde{N}}$.  Thus, when $\tilde{N}_{\rm init}\lesssim
\tilde{N}_{\rm eq}$, $\tilde{N}$ decays in the $\gamma_\chi$-dominated
universe and hence
\begin{eqnarray}
f_{\gamma_{\tilde{N}}}\sim
\frac
{(\tilde{N}_{\rm init}/\tilde{N}_{\rm eq})^2}
{1+(\tilde{N}_{\rm init}/\tilde{N}_{\rm eq})^2}
~~~:~~~ \Gamma_N \lesssim \Gamma_\chi,~
\tilde{N}_{\rm init} \lesssim \tilde{N}_{\rm eq}.
\label{f(init<eq)}
\end{eqnarray}
On the contrary, if $\tilde{N}_{\rm init}\gtrsim \tilde{N}_{\rm eq}$,
the right-handed sneutrino decays after it dominates the universe and
we obtain
\begin{eqnarray}
f_{\gamma_{\tilde{N}}}\sim
\frac
{(\tilde{N}_{\rm init}/\tilde{N}_{\rm eq})^{8/3}}
{1+(\tilde{N}_{\rm init}/\tilde{N}_{\rm eq})^{8/3}}
~~~:~~~ \Gamma_N\lesssim \Gamma_\chi,~
\tilde{N}_{\rm eq} \lesssim \tilde{N}_{\rm init}.
\label{f(eq<init)}
\end{eqnarray}

When $\tilde{N}$ decays, lepton-number asymmetry is also generated.
Such a lepton-number asymmetry is converted to the baryon-number
asymmetry due to the sphaleron process.  The resultant baryon-number
asymmetry depends on whether $\tilde{N}$ decays before or after the
$\tilde{N}$-dominated universe is realized.  If $\tilde{N}$ decays
later than the inflaton decay, the baryon-to-entropy ratio is given by
\cite{PRD65-043512}
\begin{eqnarray}
    \frac{n_B}{s} \simeq 
    0.24 \times 10^{-10} f_{\gamma_{\tilde{N}}} \delta_{\rm eff}
    \left(
        \frac{T_{N}}{10^6 {\rm GeV}}
    \right)
    \left(
        \frac{m_{\nu_3}}{0.05 {\rm eV}}
    \right)
    ~~~:~~~ \Gamma_N < \Gamma_\chi,
    \label{nB/s(N-dom)}
\end{eqnarray}
where $T_{N}$ is the temperature at the epoch of the decay of
$\tilde{N}$, $m_{\nu_3}$ the mass of the heaviest (left-handed)
neutrino mass.  (So, if $\tilde{N}$ decays after dominating the
universe, $T_{N}$ becomes the reheat temperature due to the decay of
$\tilde{N}$.)  In addition, in the basis where the Majorana mass
matrix for the right-handed neutrinos $\hat{M}$ is real and
diagonalized, the effective CP violating phase is given by
\begin{eqnarray}
    \delta_{\rm eff} = \frac{\langle H_u\rangle^2}{m_{\nu_3}}
    \frac{ {\rm Im} 
    [ \hat{h}\hat{h}^\dagger \hat{M}^{-1}\hat{h}^*\hat{h}^T ]_{11} }
    { [\hat{h} \hat{h}^\dagger]_{11} },
\end{eqnarray}
where $\hat{h}_{i\alpha}$ is the neutrino Yukawa matrix with $i$ and
$\alpha$ being the generation indices for the right-handed and
left-handed neutrinos, respectively.  Notice that, with a maximum CP
violation, $\delta_{\rm eff}\sim 1$.  On the contrary, if
$\Gamma_\chi<\Gamma_N$, $\tilde{N}$ decays before the decay of the
inflaton field.  In this case, radiation-dominated universe is
realized much later than the epoch of the sneutrino decay.  In this
case, the baryon-to-entropy ratio is given by
\begin{eqnarray}
    \frac{n_B}{s} \simeq
    0.24 \times 10^{-10} \delta_{\rm eff}
    \left(
        \frac{T_{\rm R}}{10^6 {\rm GeV}}
    \right)
    \left(
        \frac{m_{\nu_3}}{0.05 {\rm eV}}
    \right)
    \left( \frac{\tilde{N}_{\rm init}}{M_*} \right)^2
    ~~~:~~~ \Gamma_\chi < \Gamma_N,
    \label{nB/s(N-nondom)}
\end{eqnarray}
where $T_{\rm R}$ is the reheating temperature due to the decay of the
inflaton field.  

As one can see, large enough baryon asymmetry ($n_B/s\sim 8\times
10^{-11}$ \cite{aph0205931}) can be generated in this scenario with
relatively low reheating temperature.  In particular, even if we
impose the upper bound on the reheating temperature to avoid the
overproduction of the gravitinos, it is possible to generate large
enough baryon asymmetry.  For $\Gamma_N< \Gamma_\chi$, requiring
$T_N\lesssim 10^{9-10}\ {\rm GeV}$ in order not to overproduce
gravitinos \cite{gravitino}, $n_B/s\sim 10^{-10}$ can be realized when
$f_{\gamma_{\tilde{N}}}\gtrsim 10^{-3} - 10^{-2}$.  For $\Gamma_\chi<
\Gamma_N$, much smaller $f_{\gamma_{\tilde{N}}}$ is possible to
generate enough baryon asymmetry even if we require $T_{\rm R}\lesssim
10^{9-10}\ {\rm GeV}$ to avoid the gravitino problem.  In addition, it
is notable that $n_B/s\sim 10^{-10}$ can be realized even with
$f_{\gamma_{\tilde{N}}}\ll 1$ if we push up $T_N$ or $T_{\rm R}$.

\section{Density Perturbations}
\label{sec:perturbations}
\setcounter{equation}{0}

As we have seen, in the scenario we consider, baryon asymmetry as well
as some fraction of the CMB radiation are generated from the decay
product of $\tilde{N}$.  Thus, if $\tilde{N}$ has a primordial
fluctuation, such a fluctuation also becomes the source of the cosmic
density perturbations.  Most importantly, such a fluctuation affects
the CMB anisotropy which is now being measured very precisely with
various experiments
\cite{APJ464L1,aph0104460,aph0104459,aph0104489,MAP}.  

Fluctuation of $\tilde{N}$ is primarily induced during the inflation;
assuming that the (effective) mass of the right-handed sneutrino
during the inflation is smaller than the expansion rate during the
inflation $H_{\rm inf}$, the primordial fluctuation of $\tilde{N}$ is
estimated as
\begin{eqnarray}
\delta \tilde{N}_{\rm init} = \frac{H_{\rm inf}}{2\pi}.
\label{dN_init}
\end{eqnarray}

If $\delta\tilde{N}_{\rm init}$ is non-vanishing, this becomes a
source of the entropy between components generated from the decay
products of $\chi$ and $\tilde{N}$.  It is convenient to define the
parameter
\begin{eqnarray}
S_{\tilde{N}\chi}^{(\delta\tilde{N})} \equiv 
\delta_{\tilde{N}}^{(\delta\tilde{N})} - 
\delta_\chi^{(\delta\tilde{N})},
\label{S_init}
\end{eqnarray}
where $\delta_X\equiv\delta\rho_X /\rho_X$ with $\rho_X$ being the
energy density of the component $X$ and $\delta\rho_X$ its
perturbation in the Newtonian gauge.  Here, the right-hand side is
evaluated after $\tilde{N}$ and $\chi$ both start to oscillate, and
the superscript ``$(\delta\tilde{N})$'' is for variables generated
from the primordial fluctuation of $\tilde{N}$.  Solving the equations
of motions for the scalar fields, we obtain
\cite{PLB522-215,PRD66-063501}
\begin{eqnarray}
S_{\tilde{N}\chi}^{(\delta\tilde{N})} = 
\frac{2\delta\tilde{N}_{\rm init}}{\tilde{N}_{\rm init}},
\end{eqnarray}
where we assumed that the potential of the scalar fields are dominated
by the quadratic terms.  After the decays of $\tilde{N}$ and $\chi$,
$S_{\tilde{N}\chi}^{(\delta\tilde{N})}$ becomes entropy between
components generated from the decay products of $\tilde{N}$ and $\chi$
for the perturbations with wavelength much longer than the horizon
scale.

With the non-vanishing value of the primordial fluctuation of
$\tilde{N}$ given in Eq.\ (\ref{dN_init}), it is important to note
that there are two independent sources of the perturbations in our
case; fluctuation of the inflaton field and that of the right-handed
sneutrino.  In the framework of the linear perturbation theory, we can
discuss effects of these perturbations separately.  In addition, in
discussing the CMB angular power spectrum $C_l$, effects of the two
perturbations can be treated separately since we assume no correlation
between these two fields.  Thus, the total angular power spectrum is
given in the form
\begin{eqnarray}
C_l = C_l^{(\delta\chi)} + C_l^{(\delta\tilde{N})},
\label{Cl(tot)}
\end{eqnarray}
where $C_l^{(\delta\chi)}$ and $C_l^{(\delta\tilde{N})}$ are the
contributions from the primordial fluctuations in the inflaton and
$\tilde{N}$, respectively.  The inflaton contribution
$C_l^{(\delta\chi)}$ is known to become the adiabatic result (with
relevant spectral power index).  Thus, let us consider the second term
$C_l^{(\delta\tilde{N})}$.

In discussing the cosmic density perturbations induced from
$\delta\tilde{N}_{\rm init}$, we treat the radiations originated from
these fields separately; $\gamma_\chi$ for the radiation from the
inflaton and $\gamma_{\tilde{N}}$ which is that from the right-handed
sneutrino.  Density perturbations for these components are defined
separately:
$\delta_{\gamma_X}\equiv\delta\rho_{\gamma_X}/\rho_{\gamma_X}$ and
$V_{\gamma_X}$ are density and velocity fluctuations of $\gamma_X$
with $X$ being $\chi$ and $\tilde{N}$.  (Hereafter, we use the
Newtonian gauge.  We follow the notation and convention of
\cite{HuThesis}.)

In the following, we follow the evolution of the density perturbations
of various components solving the relevant Einstein and Boltzmann
equations.  In particular, we are interested in evolutions of the
density perturbations when $H\gtrsim \Gamma_N$ and properties of the
density perturbations just after the decay of $\tilde{N}$.  In this
case, we consider the evolution of the perturbations in the universe
with a very high temperature where various charged particles are
thermally produced.  Then, the radiation component becomes locally
isotropic and hence the anisotropic stress perturbation can be
neglected.

Denoting the perturbed line element as
\begin{eqnarray}
    ds^2 = a^2 
    \left[ - (1 + 2\Psi) d\tau^2 + (1 + 2\Phi) 
    \delta_{ij} dx^i dx^j \right],
    \label{ds}
\end{eqnarray}
with $\tau$ being the conformal time and $a$ being the scale factor,
the Poisson equation for the metric perturbation is given by
\begin{eqnarray}
k^2 \Phi
= 4 \pi G a^2 \rho_{\rm T}
\left[ \delta_{\rm T} 
+ 3 \frac{a'}{a} (1 + w_{\rm T}) V_{\rm T}/k \right],
\label{PoissonEq}
\end{eqnarray}
where $k$ is the comoving momentum, $G$ the Newton constant, and the
``prime'' denotes the derivative with respect to the conformal time
$\tau$.  In addition, the subscript ``T'' is for the total matter and
$w_{\rm T}$ denotes the equation-of-state parameter of the total
matter.  On the other hand, neglecting the contribution from the
anisotropic stress tensor, evolution of the density and velocity
perturbations in the Fourier space are given by \cite{HuThesis}
\begin{eqnarray}
\delta_{\gamma_X}' = -\frac{4}{3} k V_{\gamma_X} - 4\Phi',~~~
V_{\gamma_X}' = k \left( \frac{1}{4} \delta_{\gamma_X} + \Psi
\right).
\label{d'&V'}
\end{eqnarray}
In addition, in the situation we consider, the relation $\Phi=-\Psi$
holds since the anisotropic stress tensor is small enough.  In the
following, we use this relation to eliminate $\Phi$.

Using Eqs.\ (\ref{PoissonEq}) and (\ref{d'&V'}), we can discuss the
evolution of the cosmic density perturbations induced from the
primordial fluctuation of the amplitude of the right-handed sneutrino
condensation.  In this case, we neglect the adiabatic fluctuations
generated by the inflaton fluctuation.  In addition, since we are
interested in the perturbations whose wavelength is much longer than
the horizon scale (at the epoch of the baryogenesis and reheating), we
can expand the solution as a function of $k\tau$.

Evolution of $\delta_{\gamma_\chi}$ is quite simple.  Just after the
inflation, the right-handed sneutrino is the sub-dominant component
and hence $\Psi^{(\delta\tilde{N})}$ vanishes as $\tau\rightarrow 0$.
In addition, since there is no primordial perturbation in the inflaton
sector for this mode,
$\delta_{\gamma_\chi}^{(\delta\tilde{N})}\rightarrow 0$ as
$\tau\rightarrow 0$.  Thus, we obtain
\begin{eqnarray}
\delta_{\gamma_\chi}^{(\delta\tilde{N})} = 
4\Psi^{(\delta\tilde{N})} + O(k^2\tau^2).
\end{eqnarray}
This relation holds at any moment of the evolution of the universe.

Using the above relation, the entropy between the radiation and the
baryon components can be evaluated.  We calculate the entropy in the
radiation-dominated universe after the decay of both $\chi$ and
$\tilde{N}$.  In the radiation-dominated universe \cite{HuThesis}
\begin{eqnarray}
\Delta_{\rm T} = O(k^2 \tau^2),~~~
\delta_{\rm T} = -2 \Psi_{\rm RD},~~~
V_{\rm T} = \frac{1}{2} \Psi_{\rm RD} k \tau.
\end{eqnarray}
(Here and hereafter, we only denote the leading contribution to the
perturbations.)  Then, using the relation
\begin{eqnarray}
\Delta_{\rm T} = f_{\gamma_{\chi}} \Delta_{\gamma_{\chi}} +
f_{\gamma_{\tilde{N}}} \Delta_{\gamma_{\tilde{N}}},
\end{eqnarray}
with $\Delta_{\gamma_X}=\delta_{\gamma_X}+4 (a'/a)V_{\rm T}/k$, we
obtain
\begin{eqnarray}
\Delta_{\gamma_{\tilde{N}}}^{(\delta\tilde{N})} =
- \frac{f_{\gamma_{\chi}}}{f_{\gamma_{\tilde{N}}}}
\Delta_{\gamma_{\chi}}^{(\delta\tilde{N})} =
- 6\frac{f_{\gamma_{\chi}}}{f_{\gamma_{\tilde{N}}}}
\Psi_{\rm RD}^{(\delta\tilde{N})}.
\end{eqnarray}

Using the fact that the entropy between any component produced from
$\tilde{N}$ and that from the inflaton field is conserved, we can
relate $\Psi_{\rm RD}^{(\delta\tilde{N})}$ with
$S_{\tilde{N}\chi}^{(\delta\tilde{N})}$; with relation
$S_{\tilde{N}\chi}^{(\delta\tilde{N})}
=\frac{3}{4}(\Delta_{\gamma_{\tilde{N}}}^{(\delta\tilde{N})} -
\Delta_{\gamma_{\chi}}^{(\delta\tilde{N})})$, we obtain
\begin{eqnarray}
\Psi_{\rm RD}^{(\delta\tilde{N})} = 
-\frac{2}{9} f_{\gamma_{\tilde{N}}} 
S_{\tilde{N}\chi}^{(\delta\tilde{N})}
= -\frac{4}{9} f_{\gamma_{\tilde{N}}} 
\frac{\delta\tilde{N}_{\rm init}}{\tilde{N}_{\rm init}}.
\end{eqnarray}
Thus, if $\tilde{N}$ decays after it dominates the universe,
$f_{\gamma_{\tilde{N}}}\simeq 1$ and hence the metric perturbation
becomes comparable to the primordial entropy perturbation.  On the
other hand, if $f_{\gamma_{\tilde{N}}}\ll 1$, the metric perturbation
becomes negligibly small.

Since the baryon asymmetry is generated from $\tilde{N}$, the density
fluctuation in the baryonic component is given by
\begin{eqnarray}
\Delta_b^{(\delta\tilde{N})} = 
\frac{3}{4} \Delta_{\gamma_{\tilde{N}}}^{(\delta\tilde{N})} = 
- \frac{9}{2} \frac{f_{\gamma_{\chi}}}
{f_{\gamma_{\tilde{N}}}} \Psi_{\rm RD}^{(\delta\tilde{N})}.
\end{eqnarray}
Thus, the entropy between the baryon and the radiation is given by
\begin{eqnarray}
S_{b\gamma}^{(\delta\tilde{N})} = 
\Delta_b^{(\delta\tilde{N})} - 
\frac{3}{4} \Delta_{\rm T}^{(\delta\tilde{N})}
= - \frac{9}{2} 
\frac{f_{\gamma_{\chi}}}{f_{\gamma_{\tilde{N}}}} 
\Psi_{\rm RD}^{(\delta\tilde{N})}
= -
\frac{9(1 - f_{\gamma_{\tilde{N}}})}
{2f_{\gamma_{\tilde{N}}}} 
\Psi_{\rm RD}^{(\delta\tilde{N})}.
\label{S_bg}
\end{eqnarray}
Thus, the correlated mixture of the adiabatic and isocurvature
fluctuations is generated.  In addition, if the right-handed sneutrino
once dominates the universe, $f_{\gamma_{\tilde{N}}}\rightarrow 1$ and
hence the perturbation becomes adiabatic
\cite{PLB522-215,PRD66-063501,NPB626-395,PLB524-5}.  (Thus,
in this case, $\tilde{N}$ may play the role of the so-called
``curvaton'' field.)  On the contrary, if the inflaton field and its
decay products always dominate the universe,
$f_{\gamma_{\tilde{N}}}\rightarrow 0$ and the perturbation becomes
isocurvature.  Thus, the interesting spectrum would be obtained when
the inflaton and the right-handed sneutrino both produce significant
amount of the resultant radiation.

\section{CMB Angular Power Spectrum}
\label{sec:cmb}
\setcounter{equation}{0}

Now we discuss the CMB angular power spectrum induced from the
primordial fluctuation of $\tilde{N}$.  For this purpose, we
parameterize the entropy perturbation induced from
$\delta\tilde{N}_{\rm init}$ as 
\begin{eqnarray}
S_{b\gamma}^{(\delta\tilde{N})} = 
\kappa_b \Psi_{\rm RD}^{(\delta\tilde{N})}.
\end{eqnarray}
Then, using Eqs.\ (\ref{f(Gn>Gx)}), (\ref{f(init<eq)}),
(\ref{f(eq<init)}), and (\ref{S_bg}), the $\kappa_b$ parameter is
estimated as
\begin{eqnarray}
\kappa_b \sim -\frac{9}{2} \times
\left\{ \begin{array}{ll}
(\Gamma_\chi/\Gamma_N)^{-2/3}(\tilde{N}_{\rm init}/M_*)^{-2} &
~~~:~~~\Gamma_\chi \lesssim\Gamma_N
\\
(\tilde{N}_{\rm init}/\tilde{N}_{\rm eq})^{-2} &
~~~:~~~ \Gamma_N \lesssim \Gamma_\chi,~
\tilde{N}_{\rm init} \lesssim \tilde{N}_{\rm eq}
\\
(\tilde{N}_{\rm init}/\tilde{N}_{\rm eq})^{-8/3} &
~~~:~~~ \Gamma_N\lesssim \Gamma_\chi,~
\tilde{N}_{\rm eq} \lesssim \tilde{N}_{\rm init}
\end{array} \right. .
\end{eqnarray}
Thus, in our case, the $\kappa_b$ parameter varies from $-\infty$ to
$0$.  If $\tilde{N}_{\rm init}\ll\tilde{N}_{\rm eq}$, the $\kappa_b$
parameter goes to $-\infty$ and hence the density perturbation becomes
(baryonic) isocurvature.  On the contrary, when $\tilde{N}_{\rm
init}\gg\tilde{N}_{\rm eq}$, $\kappa_b\sim 0$ and the adiabatic
density perturbation is obtained from the primordial fluctuation of
$\tilde{N}$.  The most interesting case is $\tilde{N}_{\rm
init}\sim\tilde{N}_{\rm eq}$, where $\kappa_b\sim O(0.1-1)$.  In this
case, correlation between adiabatic and isocurvature perturbations
becomes the most effective.  For the case where $\Gamma_N<
\Gamma_\chi$, the requirement of enough baryon asymmetry with the
gravitino constraint give $f_{\gamma_{\tilde{N}}}\gtrsim 10^{-3} -
10^{-2}$, and hence $\kappa_b\gtrsim -1000$.  If $\Gamma_\chi<
\Gamma_N$, smaller $\kappa_b$ is possible.  (See Eqs.\ 
(\ref{nB/s(N-dom)}) and (\ref{nB/s(N-nondom)}).)

\begin{figure}[t]
\centerline{\epsfxsize=0.55\textwidth\epsfbox{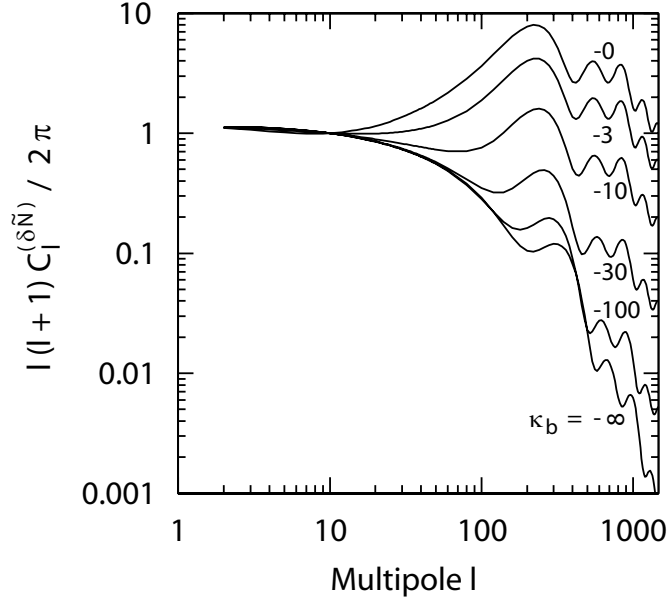}}
\caption{The CMB angular power spectrum from the primordial 
fluctuation of the right-handed sneutrino.  The $\kappa_b$ parameter
is taken as $0$, $-3$, $-10$, $-30$, $-100$, and $-\infty$ from above.
We consider the flat universe with $\Omega_bh^2=0.02$, $\Omega_m=0.3$,
and $h=0.65$, \cite{aph0205931} and the initial power spectral indices
for primordial density perturbations are all assumed to be 1 (i.e., we
adopt scale-invariant initial power spectra).  The overall
normalizations are taken as
$[l(l+1)C_l^{(\delta\tilde{N})}/2\pi]_{l=10}=1$.}
\label{fig:kappa-dep}
\end{figure}

In Fig.\ \ref{fig:kappa-dep}, we plot the CMB angular power spectrum
induced from the primordial fluctuation of the right-handed sneutrino
$C_l^{(\delta\tilde{N})}$.  (Here and hereafter, we assume there is no
entropy perturbation in the CDM sector.)  Notice that the lines with
$\kappa_b=0$ and $\kappa_b=-\infty$ coincide with the results with
purely adiabatic and isocurvature density perturbations, respectively.
For a general case, however, the angular power spectrum has a unique
structure.  In particular, as one can see, the acoustic peaks are
suppressed relative to the Sachs-Wolfe (SW) tail as $|\kappa_b|$
increases.

As we discussed, the total angular power spectrum is given by the sum
of the inflaton and the $\tilde{N}$ contributions, as shown in Eq.\
(\ref{Cl(tot)}).  Here, $C_l^{(\delta\chi)}$ is the inflaton
contribution which is parameterized by the primordial metric
perturbation generated from the inflaton perturbation $\Psi^{\rm
(\delta\chi)}$, which is given by \cite{PRD28-629}
\begin{eqnarray}
\Psi^{(\delta\chi)}_{\rm RD} = \frac{4}{9}
\left[ \frac{H_{\rm inf}}{2\pi}\frac{3H_{\rm inf}^2}{V_{\rm inf}'} 
\right]_{k=aH_{\rm inf}},
\label{Psi(dchi)_RD2}
\end{eqnarray}
where $V_{\rm inf}$ is the inflaton potential and $V_{\rm inf}'\equiv
(\partial V_{\rm inf}/\partial\chi)$, and the superscript
$(\delta\chi)$ means that the corresponding variable is generated from
the inflaton fluctuation.  On the contrary, $C_l^{(\delta\tilde{N})}$
is from the primordial fluctuation of $\tilde{N}$, and is
parameterized by $S_{\tilde{N}\chi}^{(\delta\tilde{N})}$ given in Eq.\
(\ref{S_init}).  Thus, the total angular power spectrum crucially
depends on three parameters, $\kappa_b$, $\Psi^{(\delta\chi)}$, and
$S_{\tilde{N}\chi}^{(\delta\tilde{N})}$.  To parameterize the relative
size between $\Psi^{(\delta\chi)}$ and
$S_{\tilde{N}\chi}^{(\delta\tilde{N})}$, we define
\begin{eqnarray}
R_b = S_{\tilde{N}\chi}^{(\delta\tilde{N})} / 
\Psi^{(\delta\chi)}_{\rm RD}.
\end{eqnarray}
(Notice that, with the above definition,
$\Psi^{(\delta\tilde{N})}_{\rm
RD}=-\frac{2}{9}f_{\gamma_{\tilde{N}}}R_b\Psi^{(\delta\chi)}_{\rm
RD}$.)  The $R_b$-parameter depends on the initial amplitude of the
right-handed sneutrino as well as the model of inflation.  For the
chaotic inflation model, for example, $R_b\simeq 0.6M_*/\tilde{N}_{\rm
init}$ \cite{PRD66-063501}.

\begin{figure}
\centerline{\epsfxsize=0.55\textwidth\epsfbox{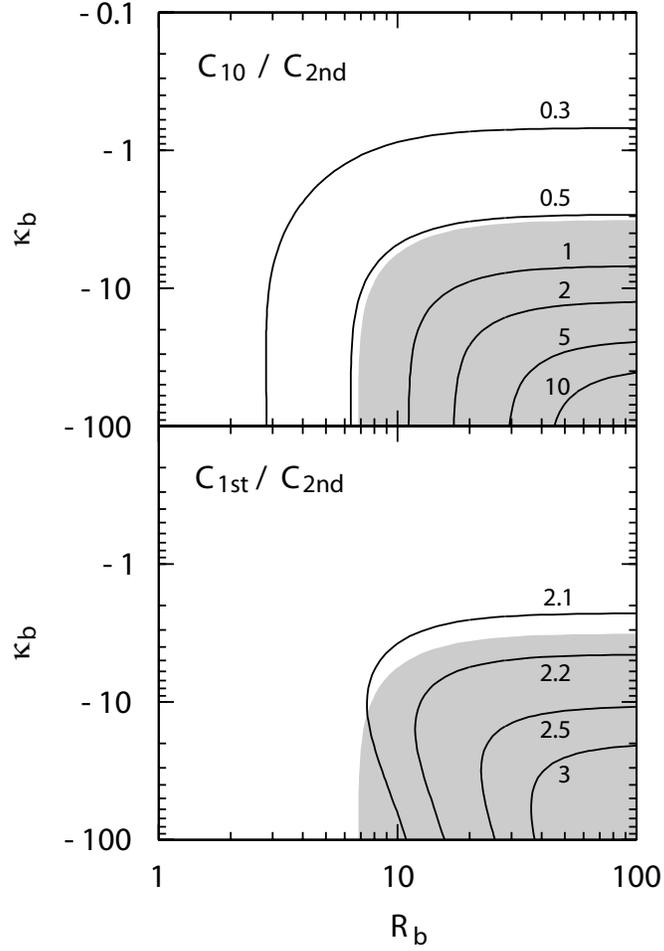}}
\caption{The ratios $C_{10}/C_{\rm 2nd}$ (top) and $C_{\rm 1st}/C_{\rm
2nd}$ (bottom) on the $R_b$ vs.\ $\kappa_b$ plane.  (The numbers in
the figure are the corresponding ratios.)  The shaded region
corresponds to the parameter space with $\chi^2>84$.  The cosmological
parameters are the same as those used in Fig.\ \ref{fig:kappa-dep}.}
\label{fig:Cl/Cl}
\end{figure}

Then, the shape of the total angular power spectrum is determined once
the parameters $R_b$ and $\kappa_b$ (as well as other cosmological
parameters) are fixed.  As these parameters vary, the shape of the
angular power spectrum changes as follows.  As the $R_b$-parameter is
increased, $C_l^{(\delta\tilde{N})}$ is more enhanced and hence the
height of the acoustic peaks are suppressed relative to the SW tail.
In addition, as $|\kappa_b|$ is increased, $C_l$ at high multipole is
suppressed relative to that at low one since the acoustic peaks of
$C_l^{(\delta\tilde{N})}$ is suppressed in this case.  In Fig.\ 
\ref{fig:Cl/Cl}, we plot the ratios $C_{10}/C_{\rm 2nd}$ and $C_{\rm
1st}/C_{\rm 2nd}$ on the $R_b$ vs.\ $\kappa_b$ plane.

As one can read off from Fig.\ \ref{fig:Cl/Cl}, the shape of the
angular power spectrum deviates from that from purely adiabatic
density perturbations if $R_b$ and $\kappa_b$ are non-vanishing.
Since the current data of the CMB angular power spectrum is well
consistent with $C_l$ calculated from the purely adiabatic density
perturbations, $C_l$ in our scenario becomes inconsistent with the
experiments if the $R_b$ and $|\kappa_b|$ become too large.  In order
to derive the constraint on these parameters, we calculate the
goodness-of-fit parameter $\chi^2=-2\ln L$, where $L$ is the
likelihood function, as a function of $R_b$ and $\kappa_b$.  In our
calculation, the offset log-normal approximation is used
\cite{APJ533-19}, and we use a data set consisting of 65 data points;
24 from COBE \cite{APJ464L1}, 19 from BOOMERanG \cite{aph0104460}, 13
from MAXIMA \cite{aph0104459}, and 9 from DASI \cite{aph0104489}.
Then, in Fig.\ \ref{fig:Cl/Cl}, we shaded the region where $\chi >84$,
which corresponds to 95 \% C.L.\ excluded region for the $\chi^2$
statistics with 64 degrees of freedom.  As expected, the angular power
spectrum at lower multipoles is enhanced relative to that at higher
multipoles.  In particular, the the ratio $C_{10}/C_{\rm 2nd}\simeq
0.25$ in the purely adiabatic case, and hence the height of the SW
tail can be enhanced by factor $\sim 2$ in the case of the sneutrino
leptogenesis.  In addition, $C_{\rm 1st}/C_{\rm 2nd}\simeq 2.0$ in the
purely adiabatic case, and hence the enhancement of the ratio $C_{\rm
1st}/C_{\rm 2nd}$ is $5\ \%$ or so in this case.

Precise determination of the CMB angular power spectrum by the MAP
experiment will provide stronger constraints on our scenario.  At the
MAP experiment, uncertainty in $C_l$ for multipoles $l\lesssim 1000$
is expected to be dominated by the cosmic variance, and in this case
the error of $C_l$ is given by \cite{PRD52-4307}
\begin{eqnarray}
\delta C_l = \sqrt{ \frac{2}{2l+1} } C_l.
\end{eqnarray}
Thus, error of single $C_l$ may not be small.  However, combining the
informations derived from $C_l$ with different $l$, uncertainties can
be reduced.  For example, if we use the data for $2\leq l\leq 50$,
$201\leq l\leq 250$, and $526\leq l\leq 575$ to estimate the heights
of the SW tail (represented by $C_{10}$) and the first and acoustic
peaks, the errors are estimated as $\delta C_{10}/C_{10}\simeq 1.5\ 
\%$, $\delta C_{\rm 1st}/C_{\rm 1st}\simeq 0.9\ \%$, and $\delta
C_{\rm 2nd}/C_{\rm 2nd}\simeq 0.6\ \%$.  With this accuracy, effects
of isocurvature fluctuations will be observed if the isocurvature
contribution enhances the height of the SW tail by a few \% or so.

One important question is whether we can distinguish the case with
correlated mixture of the adiabatic and isocurvature fluctuations from
the case with uncorrelated isocurvature fluctuation (i.e.,
$\kappa_b\rightarrow\infty$).  Since the angular power spectrum at low
multipoles are suppressed relative to $C_l$ at higher multipoles in
both cases, two cases are indistinguishable only by determining the
ratio $C_{10}/C_{\rm 2nd}$ or $C_{\rm 1st}/C_{\rm 2nd}$.  As seen in
Fig.\ \ref{fig:Cl/Cl}, however, contours of constant $C_{10}/C_{\rm
2nd}$ and those of $C_{\rm 1st}/C_{\rm 2nd}$ on the $R_b$ vs.\ 
$\kappa_b$ plane are not parallel and hence these two cases can be, in
principle, distinguished by simultaneously determining the ratios
$C_{10}/C_{\rm 2nd}$ and $C_{\rm 1st}/C_{\rm 2nd}$.

\begin{figure}[t]
\centerline{\epsfxsize=0.55\textwidth\epsfbox{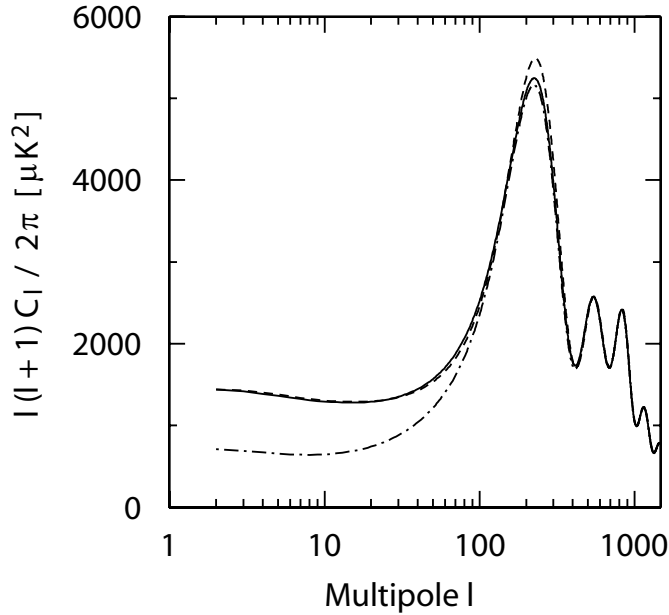}}
\caption{The total CMB angular power spectrum with
$(R_b, \kappa_b)=(100, -3)$ (dashed) and $(6.4, -100)$ (solid).  The
cosmological parameters are the same as those used in Fig.\ 
\ref{fig:kappa-dep}, and the normalizations are arbitrary.  For
comparison, we also plot the result in the purely adiabatic case
(dot-dashed).}
\label{fig:Cladiiso}
\end{figure}

To discuss this issue, in Fig.\ \ref{fig:Cladiiso}, we plot the CMB
angular power spectrum for $(R_b, \kappa_b)=(100, -3)$ (i.e., for the
case with the correlated mixture of the adiabatic and isocurvature
fluctuations) and $(6.4, -100)$ (i.e., for the case with uncorrelated
isocurvature fluctuations).  Here, we choose parameters such that the
ratio $C_{10}/C_{\rm 2nd}$ becomes the same for two cases.  Even if
the ratio $C_{10}/C_{\rm 2nd}$ does not differ, the heights of the
first acoustic peak are different; in this case, the ratio $C_{\rm
1st}/C_{\rm 2nd}$ differs about $5\ \%$, which is within the reach of
the MAP experiment if the error in $C_l$ mentioned before is realized.
Thus, if the effect of the isocurvature mode is relatively large, an
evidence of the correlated mixture of the adiabatic and isocurvature
fluctuations may be observed by the MAP experiment.

\section{Summary}
\label{sec:summary}
\setcounter{equation}{0}

In this paper, we discussed the CMB anisotropy in the scenario where
the baryon asymmetry of the universe originates in a scalar field
condensation.  We have seen that, in such a scenario, correlated
mixture of the adiabatic and isocurvature fluctuations may be
generated in particular when the decay product of the inflaton field
and that of the scalar field (i.e., $\tilde{N}$ in our example) both
significantly contribute to the present CMB radiation.  With such a
correlated mixture of fluctuations, the CMB angular power spectrum may
be significantly affected and the on-going MAP experiment may observe
the signal of the baryogenesis from the scalar-field condensation.

In our discussion, we used the sneutrino-induced leptogenesis as an
example to make our discussion clearer.  However, our discussion can
be applied to a wide class of scenarios where the baryon asymmetry of
the universe originates in a scalar-field condensation, like
Affleck-Dine scenario.

{\sl Acknowledgment:} We acknowledge the use of CMBFAST \cite{cmbfast}
and RADPACK \cite{radpack} packages for our numerical calculations.
T.M. would like to thank the theory group of the Lawrence Berkeley
National Laboratory where this work was initiated.  The work of
T.M. is supported by the Grant-in-Aid for Scientific Research from the
Ministry of Education, Science, Sports, and Culture of Japan, No.\
12047201 and No.\ 13740138.  The work of H.M. is supported in part by
the DOE Contract DE-AC03-76SF00098 and in part by the NSF grant
PHY-0098840.


\begin{thebibliography}{100}

\bibitem{PRD23-347}
    A.H. Guth, 
    Phys.\ Rev.\ {\bf D23} (1981) 347.

\bibitem{seesaw} 
    T. Yanagida, 
    in {\sl Proceedings of the Workshop on Unified Theory and 
    Baryon Number of the Universe}, 
    eds.\ O. Sawada and A. Sugamoto (KEK, 1979) p.95;
    M. Gell-Mann, P. Ramond and R. Slansky, in {\sl Supergravity},
    eds.\ P.\ van Niewwenhuizen and D.\ Freedman 
    (North Holland, Amsterdam, 1979).

\bibitem{nu-osc} 
    A. Hallin, 
    talk given at ``XXth International Conference on Neutrino Physics
    and Astrophysics;''
    M. Smy,
    talk given at ``XXth International Conference on Neutrino Physics
    and Astrophysics;''
    M. Shiozawa
    talk given at ``XXth International Conference on Neutrino Physics
    and Astrophysics.''
    (For the conference webpage, see {\tt
    http://neutrino2002.ph.tum.de/}.)

\bibitem{PRD64-115011}
    N. Arkani-Hamed, L.J. Hall, H. Murayama, D.R. Smith and N. Weiner,
    Phys.\ Rev.\ {\bf D64} (2001) 115011.

\bibitem{PRD64-053005}
    F. Borzumati and Y. Nomura,
    Phys.\ Rev.\ {\bf D64} (2001) 053005.

\bibitem{PRL84-4039}
    K. Dick, M. Lindner, M. Ratz and D. Wright,
    Phys.\ Rev.\ Lett.\ {\bf 84} (2000) 4039.

\bibitem{hph0206177}
    H. Murayama and A. Pierce,
    hep-ph/0206177.

\bibitem{PLB322-349}
    H. Murayama and T. Yanagida,
    Phys.\ Lett.\ {\bf B322} (1994) 349.

\bibitem{PRD65-043512}
    K. Hamaguchi, H. Murayama and T. Yanagida,
    Phys.\ Rev.\ {\bf D65} (2002) 043512.

\bibitem{PLB155-36}
    V.A. Kuzmin, V.A. Rubakov and M.E. Shaposhnikov,
    Phys.\ Lett.\ {\bf B155} (1985) 36.

\bibitem{PLB174-45}
    M.~Fukugita and T.~Yanagida,
    Phys.\ Lett.\ {\bf B174} (1986) 45.

\bibitem{NPB249-361}
    I. Affleck and M. Dine,
    Nucl.\ Phys.\ {\bf B249} (1985) 361.

\bibitem{PRL70-1912}
    H. Murayama, H. Suzuki, T. Yanagida and J. Yokoyama,
    Phys.\ Rev.\ Lett.\ {\bf 70} (1993) 1912.

\bibitem{aph0205931}
    J.R. Primack,
    astro-ph/0205391.

\bibitem{gravitino}
    M. Kawasaki and T. Moroi,
    Prog.\ Theor.\ Phys.\ {\bf 93} 879;
    M. Kawasaki, K. Kohri and T. Moroi,
    Phys.\ Rev.\ {\bf D63} (2001) 103502.

\bibitem{APJ464L1}
    C. Bennett et al.,
    Astrophys.\ J.\ Lett.\ {\bf 464} (1996) L1.

\bibitem{aph0104460}
    C.B.\ Netterfield et al., 
    astro-ph/0104460.

\bibitem{aph0104459}
    A.T.\ Lee et al., 
    astro-ph/0104459.

\bibitem{aph0104489}
    N.W.\ Halverson et al., 
    astro-ph/0104489.

\bibitem{MAP}
    MAP webpage,
    {\tt http://map.gsfc.nasa.gov}.

\bibitem{PLB522-215}
    T. Moroi and T. Takahashi,
    Phys.\ Lett.\ {\bf B522} (2001) 215.

\bibitem{PRD66-063501}
    T. Moroi and T. Takahashi,
    Phys.\ Rev.\ {\bf D66} (2002) 063501.

\bibitem{HuThesis}
    W. Hu,
    Ph.D thesis (astro-ph/9508126).

\bibitem{NPB626-395}
    K. Enqvist and M.S. Sloth,
    Nucl.\ Phys.\ {\bf B626} (2002) 395.

\bibitem{PLB524-5}
    D.H. Lyth and D. Wands,
    Phys.\ Lett. {\bf B524} (2002) 5.

\bibitem{PRD28-629}
    J.M. Bardeen, P.J. Steinhardt and M.S. Turner,
    Phys.\ Rev.\ {\bf D28} (1983) 629.

\bibitem{APJ533-19}
    J.R. Bond, A.H. Jaffe and L. Knox,
    Astrophys.\ J.\ {\bf 533} (2000) 19.

\bibitem{PRD52-4307}
    L. Knox,
    Phys.\ Rev.\ {\bf D52} (1995) 4307.

\bibitem{cmbfast}
    M. Zaldarriaga and U. Seljak,
    Astrophys.\ J.\ {\bf 469} (1996) 437.

\bibitem{radpack}
    RADPACK webpage,
    {\tt http://flight.uchicago.edu/knox/radpack.html}.

\end{thebibliography}
\end{document}